\documentclass[preprint2]{aastex}

\begin{document}

\title{A Measurement of Time-Averaged Aerosol Optical Depth using Air-Showers Observed in Stereo by HiRes}

\date{\today}

\author{
R.U.~Abbasi,\altaffilmark{1}
T.~Abu-Zayyad,\altaffilmark{1}
J.F.~Amann,\altaffilmark{2}
G.~Archbold,\altaffilmark{1}
R.~Atkins,\altaffilmark{1}
K.~Belov,\altaffilmark{1}
J.W.~Belz,\altaffilmark{3}
S.~BenZvi,\altaffilmark{5}
D.R.~Bergman,\altaffilmark{6}
J.H.~Boyer,\altaffilmark{4}
C.T.~Cannon,\altaffilmark{1}
Z.~Cao,\altaffilmark{1}
B.M.~Connolly,\altaffilmark{5}
Y.~Fedorova,\altaffilmark{1}
C.B.~Finley,\altaffilmark{5}
W.F.~Hanlon,\altaffilmark{1}
C.M.~Hoffman,\altaffilmark{2}
M.H.~Holzscheiter,\altaffilmark{2}
G.A.~Hughes,\altaffilmark{6}
P.~H\"{u}ntemeyer,\altaffilmark{1}
C.C.H.~Jui,\altaffilmark{1}
M.A.~Kirn,\altaffilmark{3}
B.C.~Knapp,\altaffilmark{4}
E.C.~Loh,\altaffilmark{1}
N.~Manago,\altaffilmark{7}
E.J.~Mannel,\altaffilmark{4}
K.~Martens,\altaffilmark{1}
J.A.J.~Matthews,\altaffilmark{8}
J.N.~Matthews,\altaffilmark{1}
A.~O'Neill,\altaffilmark{5}
K.~Reil,\altaffilmark{1}
M.D.~Roberts,\altaffilmark{8}
S.R.~Schnetzer,\altaffilmark{6}
M.~Seman,\altaffilmark{4}
G.~Sinnis,\altaffilmark{2}
J.D.~Smith,\altaffilmark{1}
P.~Sokolsky,\altaffilmark{1}
C.~Song,\altaffilmark{5}
R.W.~Springer,\altaffilmark{1}
B.T.~Stokes,\altaffilmark{1}
S.B.~Thomas,\altaffilmark{1}
G.B.~Thomson,\altaffilmark{6}
D.~Tupa,\altaffilmark{2}
S.~Westerhoff,\altaffilmark{5}
L.R.~Wiencke,\altaffilmark{1}
A.~Zech\altaffilmark{6}\\
(The High Resolution Fly's Eye Collaboration)\\
* Corresponding author email: wiencke@cosmic.utah.edu\\
}

\altaffiltext{1}{University of Utah,
Department of Physics and High Energy Astrophysics Institute,
Salt Lake City, UT 84112, USA.}

\altaffiltext{2}{Los Alamos National Laboratory,
Los Alamos, NM 87545, USA.}

\altaffiltext{3}{University of Montana, Department of Physics and Astronomy,
Missoula, MT 59812, USA.}

\altaffiltext{4}{Columbia University, Nevis Laboratories, Irvington, 
NY 10533, USA.}

\altaffiltext{5}{Department of Physics, Columbia University, New York,
NY 10027, USA.}

\altaffiltext{6}{Rutgers --- The State University of New Jersey,
Department of Physics and Astronomy, Piscataway, NJ 08854, USA.}

\altaffiltext{7}{University of Tokyo,
Institute for Cosmic Ray Research,
Kashiwa City, Chiba 277-8582, Japan.}

\altaffiltext{8}{University of New Mexico,
Department of Physics and Astronomy,
Albuquerque, NM 87131, USA.}

\begin{abstract}

Air fluorescence measurements of cosmic ray energy must be corrected
for attenuation of the atmosphere. In this paper we show that the
air-showers themselves can yield a measurement of the aerosol
attenuation in terms of optical depth, time-averaged over extended
periods. Although the technique lacks statistical power to make the
critical hourly measurements that only specialized active instruments
can achieve, we note the technique does not depend on absolute
calibration of the detector hardware, and requires no additional
equipment beyond the fluorescence detectors that observe the air
showers. This paper describes the technique, and presents results
based on analysis of 1258 air-showers observed in stereo by the High
Resolution Fly's Eye over a four year span.

\end{abstract}


\section{Introduction}
Fluorescence detectors use the atmosphere calorimetrically to measure
the energy deposited by extensive air-showers. Ultra-violet
fluorescence emitted by particle cascades can be observed tens of
kilometers away by a photosensitive detector when the primary cosmic
particle is above 10$^{18}$ eV. The energy of the primary particle is
measured in proportion to the total number of photons yielded by the
shower. 

Monitoring atmospheric clarity is required to calibrate for
atmospheric propagation losses of light between the shower and the
detector. Obtaining this calibration requires routine measurements by
specialized equipment, generally lasers, and LIDARS. While essential,
this equipment is challenging to construct, maintain, and calibrate,
especially in the remote deserts where fluorescence detectors are
located. Active systems are limited in their beams can not be so
bright as to swamp the fluorescence detectors and cause saturation of
the data acquisition systems. Additional methods to measure the
aerosol optical depth and cross-check these conventional measurements
can be helpful, especially when no additional equipment is needed.

The High Resolution Fly's Eye observatory (HiRes), located at Dugway,
Utah, USA features two fluorescence detector stations separated by
12.6 km. (See \cite{proto00} and \cite{B2002}.) Each station views
nearly the full azimuth. The HiRes-1 station has one ring of
telescopes that view 3.5 to 16 degrees of elevation. A second ring of
telescopes extends the elevation coverage of the HiRes-2 station to 30
degrees. Each telescope features a 3.75 m$^2$ mirror that focuses
light onto a camera of 256 photomultiplier tubes (PMTs). Each PMT
views approximately (1$^\circ$x1$^\circ$).

The atmosphere is modeled using molecular scattering and ozone
absorption as a baseline. The remaining attenuation is attributed to
aerosols. The HiRes experiment includes steerable lasers used to
measure aerosol attenuation. For more details, see \cite{A2005}.

This paper describes an independent measurement of aerosol optical
depth that uses air-showers viewed in stereo. It has the advantage
that it is insensitive to the absolute photometric calibration of the
detector hardware and the total fluorescence yield that are two of the
largest uncertainties of the fluorescence technique. In the sense that
air-showers are a natural part of the primary data sample, the
technique incurs no additional cost. Furthermore, the measurement is
made over the band of wavelengths that air-showers produce, and for
the range of distances over which HiRes measures air-showers. Nor
does this analysis require a comprehensive reconstruction of the
air-shower light profile, energy, or primary particle composition; it
is enough to reconstruct the shower axis, identify segments of the
shower viewed in common by two detectors, and apply a set of selection
criteria. 

We note that the technique has limitations. The relatively low flux
of extensive air-showers restricts the statistical power of the
technique to the measurement of one parameter, total aerosol optical
depth, averaged over years. To reduce sensitivity of the result to
details of the aerosol vertical distribution, the technique assumes
that most of the aerosol is distributed below the shower segments used
in the analysis. The assumption is supported by an analysis of laser
shots (Abassi 2005), that found that the aerosol vertical distribution
is consistent with an average scale height of about 1 km. For this
analysis we use shower segments at least 1 km above the detectors.

\begin{figure}
\epsscale{1}
\label{lbal-diagram}
\plotone{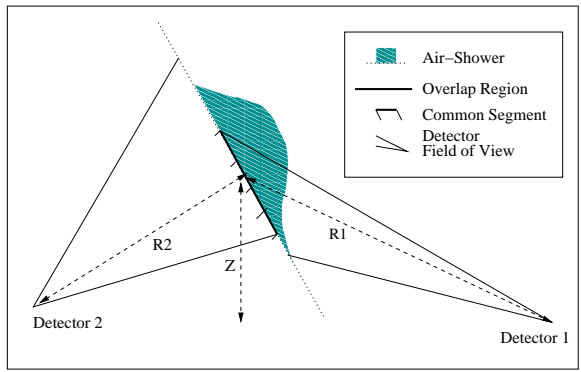}
\caption{Diagram of a cosmic ray air-shower as viewed in stereo.}
\end{figure}

\section{Stereo light balance method}
For stereo observations of atmospheric events to yield consistent results
between detectors, an accurate description of the atmospheric attenuation is
required. Conversely, a consistency constraint can be used to find the total
optical depth.

Here we use a molecular description of the atmosphere as a baseline and apply
a consistency constraint to find the remainder optical depth due to atmospheric
aerosols.

\subsection{General solution}
The aerosol atmosphere is modeled with a total aerosol optical depth
$\tau$. A ray traveling vertically to infinity is attenuated by one
exponent of $\tau$. A inclined ray traveling to an altitude $z$ is
attenuated by
\begin{equation}
T = e^{ \left( - \frac{\tau}{\sin{\alpha}} \left[ 1 - e^{ \left( -\frac{z}{SH} \right)} \right] \right)},
\end{equation}
where $T$ is transmission, $\alpha$ is the elevation angle of the ray, and $SH$
is the Scale Height of the aerosol distribution.

A useful parameter $\eta$ can be factored out of this expression.
\begin{eqnarray}
T & = & e^{ \left( - \tau \eta \right) }\\
\eta & = & \frac{1}{\sin{\alpha}} \left[ 1 - e^{ \left( -\frac{z}{SH} \right)} \right]
\end{eqnarray}
Equation 3 gives us a particular definition of $\eta$ applicable to this
model for monochromatic light in the aerosol atmosphere. However, the argument
that follows requires only that $\eta$ be a known parameter that satisfies
Equation 2.

Suppose that a segment of an extensive air-shower (Figure 1) produces
N optical photons and some number, $S_D$, are recorded by a
fluorescence detector during atmospheric conditions that are less than
perfectly clear (i.e. through aerosol haze).  Had conditions been perfectly
clear (i.e. molecular with no aerosol) a greater fraction of the photons produced
would have reached the detector.  Thus the same measured value of $S_D$
would have corresponded to a smaller number, $N_M$, of photons produced,
where $N_M<N$.

$N_M$ depends on the detected signal and, by definition, does not
depend on the aerosol property we wish to measure. It can be calculated
using $N_M=S_D*f$, where $f$ is a function of the measured
shower-detector geometry, and molecular optical depth. The latter can
be calculated from molecular scattering theory and knowledge of the
atmospheric density profile derived from radiosonde data. N and $N_M$
are related by $N=N_M/T$, ignoring multiple scattering effects.

When two detectors observe the
same shower segment, two simultaneous equations can be written.
\begin{eqnarray}
N(1) & = & N_M(1)e^{\left(\tau\eta(1)\right)} \\
N(2) & = & N_M(2)e^{\left(\tau\eta(2)\right)}
\end{eqnarray}

We constrain the two detectors to agree on the number of photons
emitted, $N(1) = N(2)$, and solve to find the light balance equation
\begin{equation}
\Delta_N = \tau \Delta_\eta, \label{lbalEqn}
\end{equation}
where 'light balance' $\Delta_N \equiv \ln{\left(\frac{N_M(1)}{N_M(2)}\right)}$
and 'event asymmetry' $\Delta_\eta \equiv \eta(2) - \eta(1)$.

It follows that a plot of $\Delta_N$ versus $\Delta_\eta$ for a sample of
events will have a slope of $\tau$.

\subsection{Polychromatic approximation}
Any application of the light balance equation (6) requires a definition of
$\eta$ which fits the attenuation model involved and satisfies Equation 2.
Equation 3 provides a definition of $\eta$ which is suitable if the
light source is monochromatic. However for polychromatic light in the
atmosphere it is not possible to satisfy Equation 2 with a simple definition of
$\eta$. This difficulty is rooted in the wavelength dependence of scattering.
Aerosol optical depth $\tau$ is approximately inversely proportional to
wavelength. Molecular scattering follows a much steeper relationship.

For convenience, we will refer to the aerosol optical depth at 355 nm as
$\tau \equiv \tau(355)$, with the understanding that depths at other
wavelengths can be scaled from this value.

A good approximate solution is found by making a guess $\tau'$ close to $\tau$
and redefining
\begin{equation}
\eta = -\frac{\ln{\left( \frac{N_M}{N'} \right)}} {\tau'},
\end{equation}
where $N'$ is the number of photons emitted by the event assuming $\tau'$, and
$N_M$ is the number of photons emitted assuming a molecular atmosphere (not
including $\tau'$).

This definition rigorously satisfies Equation 2 only if $\tau' = \tau$, and
in general it may be necessary to apply this solution iteratively to converge
on a value of $\tau$, unless the approximation is particularly good. The
quality of the approximation depends primarily on the spectral bandwidth.

In the study that follows, the sensitivity of $\tau$ with respect to $\tau'$
is less than 1:100. We will set $\tau'$ to $0.040$ for the remainder of the
discussion, since this uncertainty is much smaller than other errors in the
analysis.

\subsection{Line sources}
Equation 6 applies to point sources of light in a straightforward fashion.
An air-shower has a cross section hundreds of meters wide and is observed more
than 10 km away, and can be considered a point source traveling at the speed of
light. A simpler approach in practice, is to treat the air-shower as a line
source.

A line source can be treated by integrating an infinite number of point sources
along the line segment. Equation 6 can be applied this way, provided that
$\Delta_\eta$ is relatively constant over the segment.

In this analysis, air-shower tracks are split until $\Delta_\eta$ varies by
0.3 or less over the track segments. Detector pixel size is sometimes a
limitation in splitting the tracks. If the variation in $\Delta_\eta$ can't be
kept below 1.0, the event is removed from the data.

\subsection{Data Selection}
Data from the HiRes detectors is matched by trigger time to produce
stereo candidates. These candidates are passed through a Rayleigh
filter to select track-like events while removing various noise
triggers. Candidates may also be cut if a shower-detector plane can
not be fit. We start this analysis with 5217 stereo shower candidates
collected between December 1999 and December 2003.

\begin{deluxetable}{lccccc}
\tabletypesize{\scriptsize}
\tablecaption{Selection of real and simulated stereo data.}
\tablewidth{0pt}
\tablehead{
\colhead{Filter} & \colhead{Events} & \colhead{Segments} & \colhead{MC events} & \colhead{MC segments}
}
\startdata
0. Starting sample (see text). &
5217 & - & 9474 & -\\
1. Require commonly viewed segment(s) & 
2219 & 5492 & 5095 & 12174\\
2. High gain channel is not saturated. &
1876 & 4440 & 4211 & 9694\\
3. Rayleigh filter &
1810 & 4290 & 4199 & 9674\\
4. Track trajectory is downward. &
1803 & 4276 & 4162 & 9619\\
5. Track length is 4 degrees in each detector. &
1738 & 4211 & 3979 & 9436\\
6. Stereo plane opening angle is 8 - 172 degrees. &
1683 & 4089 & 3834 & 9090\\
7. The viewing angle is between 30 and 165 degrees. &
1572 & 3413 & 3458 & 7222\\
8. The viewing angle asymmetry is less than 50 degrees. &
1472 & 3200 & 3230 & 6625\\
9. Altitude of segment is above 1000 m. &
1362 & 2656 & 2929 & 5462\\
10. Segment spans less than 1.0 in $\Delta_\eta$. &
1258 & 2503 & 2696 & 5142\\
\enddata
\end{deluxetable}

From the stereo candidates we select 1258 events for light balance
analysis that have well reconstructed geometries and a common region
observed by both detectors. Depending on the length of the common
region, it may be divided in to segments. Tabulations for real and
simulated data are provided in Table 1 and the selection criteria is
described below. \\
1) Sometimes the two detectors view different
segments of track. These events must be cut, since there is no
overlapping region.\\
2) Events that saturate the high gain FADC channels at HiRes-2 are
dropped.\\
3) A random walk model is used to remove noise events.\\
4) The reconstructed trajectory is required to be downward.\\
5) If a track is very short, there is large uncertainty in the shower-detector
plane and therefore a potentially large uncertainty in stereo
geometry. Tracks are required to cover at least 4 degrees in each
detector.\\
6) If the opening angle is small between the two
shower-detector planes, then there is large uncertainty in the
intersection. Events with plane angles less than 8 or larger than 172
degrees are cut.\\
7) Asymmetric Cherenkov scattering is a concern when
the track is viewed at an oblique angle. Viewing angles below 30 or
above 165 degrees are cut.\\
8) To minimize the effects of any potential asymmetries, the maximum
difference in viewing angle is set at 50 degrees.\\
9) To reduce Cherenkov contribution and place observations above most aerosol, the
segment altitude must be greater than 1000 m above detectors.\\
10) Equation \ref{lbalEqn} can be applied to a linear track provided that
$\Delta_\eta$ is approximately constant over the track. Track segments are
cut if the variance in $\Delta_\eta$ across their length is greater than 1.0.
\section{Systematic Error}
Systematic errors in this analysis can not result from calibration
uncertainties of the detector hardware in the following sense. A
wavelength independent calibration scalar, $k$, applied to Equation
(5a).
\begin{mathletters}
\begin{eqnarray}
\ln{\left( k\times \frac{N(m)_1}{N(m)_2} \right)} & = & \tau \left( \eta_2-\eta_1 \right)\\
\ln{\left( \frac{N(m)_1}{N(m)_2} \right)} & = & \tau \left( \eta_2-\eta_1 \right)-\ln{(k)}
\end{eqnarray}
\end{mathletters}
becomes an additive constant ($\ln{(k)}$). This error would shift the
points in Figure 2 up or down, but would not alter the slope
($\tau$). $k$ could represent an error in overall gain in one or
both HiRes detectors, for example. A time dependent shift in
calibration could smear the data vertically thus reducing the
sensitivity of the slope measurement.

Systematic errors can arise from effects that correlate with the
aerosol optical path asymmetry $\Delta_\eta$. In this regard, we have
examined the sensitivity of the slope measurement to a number of
sources of uncertainty. Their sum in quadrature is 0.014 (see Table
2).

The aerosol vertical distribution is modeled with a 1.0 km scale
height, motivated by HiRes laser measurements \cite{A2005}. A variation in the
average scale height by $\pm0.3$ km shifts the value by $\pm0.008$. 
To estimate the effect of the Cherenkov light on the measurement, we
generated a sample of simulated showers without the Cherenkov
component. $\tau$ changed by 0.009. 

A number of other effects were also investigated. To estimate
sensitivity to wavelength dependence effects in detector response, the
shower data was reanalyzed with the calibration scaled by $\pm10\%$
per 100 nm. The difference in $\tau$ was found to be
$\mp0.002$. The analysis and simulation used a fluorescence spectrum
derived from the measurements of \cite{K1996a} and the compilation of
\cite{B1967}. Using the more recent spectrum of \cite{N2004} shifts
$\tau$ by 0.002. Uncertainty in the shower axis geometry arising
from the shower-detector plane resolution contributes an error of less
than 0.002 to $\tau$. To estimate effects of light transmission via
atmospheric multiple scattering, the data was reanalyzed with an
estimated contribution to each shower using the formalism of
\cite{R2004}. The shift in $\tau$ was 0.003. Finally, we include an 
estimated error of 0.005 that arises from the non-linear response of an
older model preamp used in some of the HiRes1 mirrors.

\begin{deluxetable}{lcc}
\tabletypesize{\scriptsize}
\tablecaption{Systematic error estimates.}
\tablewidth{0pt}
\tablehead{
\colhead{Effect} & \colhead{Approximate error to $\tau$}
}
\startdata
Cerenkov contribution & 0.009\\
Vertical aerosol distribution & 0.008\\
Preamp non-linearity & 0.005\\ 
Multiple scattering & 0.003\\
Detector wavelength dependence & 0.002\\
Geometric reconstruction & 0.002\\
Fluorescence spectrum & 0.002\\
\hline
Quadrature sum & 0.014\\
\enddata
\end{deluxetable}

\section{Results}
Figure 2 shows plots of $\Delta_N$ versus $\Delta_{\eta}$ for real data and
three Monte Carlo sets. Each point corresponds to a segment of track.

\begin{figure}
\epsscale{1}
\label{lbal-sample}
\plotone{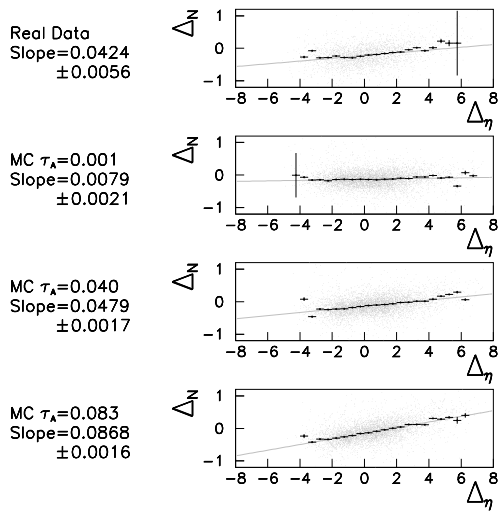}
\caption{Plots of $\Delta_N$ (light balance)
versus $\Delta_\eta$ (aerosol optical path asymmetry) for real data, and three Monte
Carlo sets. According to equation \ref{lbalEqn}, the slope of each plot should
be equal to $\tau$. The MC sets are generated using a constant $\tau$ of
0.0010, 0.0400, and 0.0833. Linear fits are made to a profile histogram, shown
in black.}
\end{figure}

The data is binned in $\Delta_\eta$. Each bin is fit to a Gaussian
to determine a mean and $\sigma$ in $\Delta_N$. The mean values each
bin are weighted by the number of entries and fit to a straight line.
Statistical uncertainties are quoted with the slopes.

Monte Carlo data is generated using random geometries and primary
particle energies, with lower energies weighted to approximate the
measured HiRes energy spectrum. Three simulated samples are generated
using three different values of $\tau$. These are listed with the
results in Figure 2. Identical reconstruction and analysis are
applied to real and simulated data samples. 

The resulting fit for the real data yields an average $\tau$ of
($0.042\pm0.006(stat)\pm0.014(sys)$)

\subsection{Cross check}
It is simple to check the accuracy of an atmospheric model by plotting light
balance as a function of path difference 
($\Delta_r=r(2)-r(1)$). (see diagram in Figure 1.) If the atmospheric model is accurate, the
slope should be zero. Figure 3 shows real data reconstructed with three model
atmospheres ($\tau$ = 0.001, 0.040, and 0.100). A positive slope will
indicate a deficit in the an optical depth of the model, while a negative slope
indicates an excess.

\begin{figure}
\epsscale{1}
\label{lbal-check}
\plotone{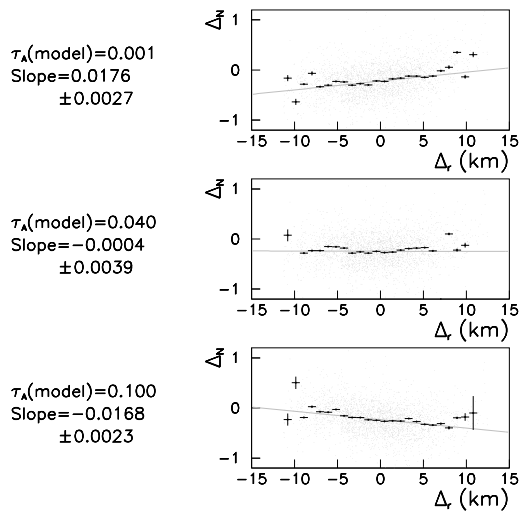}
\caption{Plots of $\Delta_N$ (light balance)
versus $\Delta_r$ (path difference) for real data using three different model
atmospheres ($\tau$ = 0.001, 0.040, and 0.100). The model with $\tau$ = 0.040
results in the smallest slope.}
\end{figure}

The plot with $\tau$ = 0.04 has a small positive slope, indicating a $\tau$
somewhat larger than 0.04, which is consistent with 0.042 shown in Figure 2.

\section{Conclusion} Stereo cosmic ray showers are used to measure
$\tau$ in a manner that is independent of absolute detector
calibration. While it cannot replace the hourly and daily
measurements obtained by specialized equipment, this method gives a
cross check of the amount of aerosols present as averaged over
extended periods. This technique may be of use to other experiments
that measure air-showers with more than one fluorescence detector
station. A trade-off between sensitivity and statistics is expected,
depending on the distance between stations. We note, in passing, that
this work is the first reported systematic use of air-showers to
estimate atmospheric clarity.

\end{document}